# Is This the Right Time to Post My Task? An Empirical Analysis on a Task Similarity Arrival in TopCoder

Razieh Saremi, Mostaan Lotfalian Saremi, Prasad Desai, and Robert Anzalone

Stevens Institute of Technology, Hoboken NJ 070390, USA
{rsaremi,mlotfali, pdesai9,ranzalon}@stevens.edu

**Abstract.** Existed studies have shown that crowd workers are more interested in taking similar tasks in terms of context, field and required technology, rather than tasks from the same project. Therefore, it is important for task owners to not only be able to plan "when the new task should arrive?" but also, to justify "what the strategic task arrival plan should be?" in order to receive a valid submission for the posted task. To address these questions this research reports an empirical analysis on the impact of similar task arrival in the platform, on both tasks' success level and workers' performance. Our study supports that 1) higher number of arrival tasks with similarity level greater than 70% will negatively impact on task competition level, 2) Bigger pool of similar open and arrival tasks would lead to lower worker attraction and elasticity, and 3) Workers who register for tasks with lower similarity level are more reliable to make a valid submission and 4) arriving task to the pool of 60% similar task will provide the highest chance of receiving valid submission.

## 1 Introduction

Crowdsourcing Software Development (CSD) requires decomposing a project to mini-tasks to put an open call for crowd workers [1,2]. This fact raises two main questions for a project manager: 1- what is the best time to crowdsource a mini-task? and 2- how can I attract skillful workers to work on my mini task?

To answer project managers' questions, a good understanding of task characteristics, task arrival, and crowd workers' sensitivity to arrival tasks are required. Apart from CSD, crowdsourcing tasks are short, simple, repetitive, requires little time and effort [3]. While in CSD, tasks are more complex, interdependent heterogamous, and requires a significant amount of time, effort [4], and expertise to achieve the task requirements. Intuitively, higher demand for skilled workers effects on their availability and increase the task failure chances.

For example, in Topcoder [5], a well-known Crowdsourcing Software platform, on average 13 tasks arrive daily added to on average 200 existing tasks, simply more demand.

Moreover, there is on average 137 active workers to take the tasks at that period which leads to on average 25 failed tasks. According to this example, there will be a long queue of tasks waiting to be taken. Considering the fixed submission date, such waiting line may result is starved tasks. Thus, task arrival policies will be one of the most important factors to avoid waiting time by assuring that there is enough available workers to take the task.

It is reported that crowd workers usually choose to register, work, and submit for tasks based on some personal utility algorithm, their skillsets and some unknown factors [6]. Crowd workers rather continue to work on similar context tasks based on their previous experience [7], task contexts include required technology, platform, and task type. Also, it is reported that one of the attractive attributes for a worker choosing a task is the monetary prize [8,9]. By arriving a higher number of tasks, crowd workers will have a higher number of different choices for taking tasks. Therefore, a higher chance of task starvation or cancelation due to zero task registration or task submission form workers may occur. This fact creates a need for a similarity algorithm to cover all mentioned utility factors for a worker to take a task. We aimed to analyze task arrival in the platform and workers' sensitivity based on similar arrival tasks in the platform in order to minimize task waiting time and task failure.

Understanding the impact of similar available tasks on arrival tasks and available workers' performance and sensitivity becomes extremely important. There is a lack of study on the impact of open similar task on the new arrival tasks in the crowdsourcing market and workers' availability and performance in the field of software crowdsourcing. Considering the schedule reduction in crowdsourcing [10], software managers are more concerned about the risks of project success. In this study, we aim at approaching these gaps by investigating the following questions:

(i) How does the number of available similar tasks impact on task failure?
(ii) How do available similar tasks effect on task competition level?
(iii) How does the queue of available similar tasks impact on task stability?

We report the design and analysis results of an empirical study based on data gathered from Topcoder, the largest software development crowdsourcing platform with an online community of over 1M crowd software workers [9].

The rest of the paper is organized as follows: Section 2 introduces the background and related work; Section 3 presents the design of the research conducted; Section 4 reports the empirical results to answer the three stated research questions. Section 5 discusses the results; and finally, Section 6 gives a summary and outlook to future work.

## 2  Background and Related Work

### 2.1  Task Similarity in CSD

Generally, workers tend to optimize their personal utility factor to register for a task [6]. It is reported that workers are more interested in working in similar tasks in terms of monetary prize [8], context and technology [7], and complexity level. Context switch generates reduction in workers' efficiency [7]. Besides the fact that workers usually register for a greater number of tasks than they can complete [11]. Combination of these two observations may lead to receiving task failure due to:
- Receiving zero registration for task based on low degree of similar tasks and lack of available skillful worker [8], and
- Receiving non-qualified submissions or zero submissions based on lack of time to work on all the registered tasks by the worker [14].

### 2.2 Workers Behavior in CSD

Software workers' arrival in the platform and their pattern of taking tasks to completion are the essential elements to shape the worker supply and demand in crowdsourcing platforms. For beginners, it takes time to improve and turn into an active worker after their first arrival [5,12]. Therefore, most of them focus on registering and gaining experience for similar tasks. Existing studies show that by passing time, registrants gaining more experience, hence better performance is expected, and consequently, valid submissions is made [13, 14]. Yet there is a small portion of workers to manage not only to make a submission but also the submission passes the peer review and mark as a valid submission [15].

A typical issue related to workers is that some workers may decide to drop certain tasks after registering for competition or possibly become inactive due to various reasons, such as different time zones and geographical distributions, different native languages spoken by software workers and number of open tasks in the workers' list of tasks [16,17,18].

Generally, workers tend to optimize their personal utility factor to register for a task [6]. It is reported that workers are more interested in working in similar tasks in terms of monetary prize [8], context and technology [7], and complexity level.

### 2.3 Decision-Making in CSD

Software Online decision algorithms have a rich literature in operations research, economics, machine learning, and artificial intelligence, etc. Most of the existing work on crowdsourcing decision making is addressing problems in the general crowdsourcing markets. For example, many studies have applied machine learning techniques in learning worker quality and optimizing task assignment decisions [19], aggregating individual answers to improve quality [20,21], and worker incentives [16]. Variety of modeling choices for repeated decision making in general crowdsourcing identified including quality of work, incentives and human factors, and performance objectives [22]. While a dynamic procurement model for crowdsourcing in which workers are required to explicitly submit their preferences has been presented [23]. The queuing theory was applied in real-time crowdsourcing to predict the expected waiting time and cost of the decomposed uploaded tasks [24]. In software crowdsourcing, only a few studies have focused on decision support for the software crowdsourcing market in terms of task pricing [6,8,25,26], developer recommendations [11,27], understanding worker behaviors [6, 14, 28,29], and improve task scheduling [7,31]. However, there is no consideration of the impact of similarity among the pool of available tasks on new arrival task status.

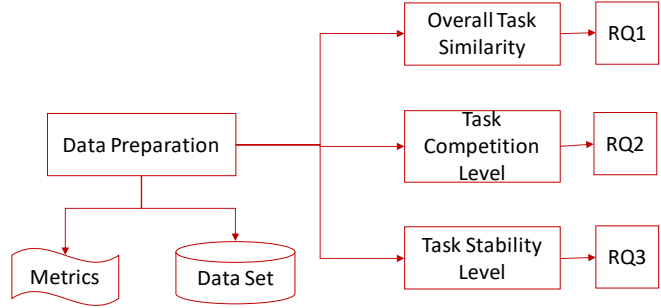

**Fig. 1.** Main flow of proposed framework and relationship to research questions

## 3 Research Design

### 3.1 Empirical Evaluation Framework

Driven by the resource-related challenges in software development, we design three evaluation studies to provide empirical evidence on the feasibility and benefits of CSD. The evaluation framework is illustrated in Fig 1.
The three research questions in this study are:
*RQ1 (Task Similarity Level):* How does the number of available similar tasks impact on task failure?
This research question aims at providing a general overview of task similarity in terms of task arrival and available similar available tasks in the platform as well as its effect on task failure rate.
*RQ2 (Task Competition Level):* How does available similar tasks effect on attracting workers?
The consistency of worker availability will be measured by comparing two different similar tasks at the same time frame. The ratio of attracting workers in the platform in the same period of time with a specific similar task will be a good measure to indicate workers' availability and attraction.
*RQ3 (Task Stability Level):* How does the queue of open tasks impact workers' performance outcomes?
The ratio of receiving valid submission per task from workers in the same period of time with a specific similar task will be a good measure to indicate task success.

### 3.2 Dataset and Metrics

The gathered dataset contains 403 individual projects including 4907 component development tasks and 8108 workers from Jan 2014 to Feb 2015, extracted from the Topcoder website [9].
Tasks are uploaded as competitions in the platform, where Crowd software workers would register and complete the challenges. On average most of the tasks have a life cycle of one and a half months from the first day of registration to the submission's deadline. When the workers submit the final files, it will be reviewed by experts to check the results and grant the scores. In order to analyze the impact of task similarity of task success in the platform, we categorized the available data and defined the following metrics, as summarized in Table 1.

**Table 1.** Summary of metrics definition

| Metric | Definition |
| --- | --- |
| Duration (D) | Total available time from registration date to submissions deadline. Range: $(0, \infty)$ |
| Task registration start date (TR) | The time when a task is available online for workers to register |
| Task submission end date (TS) | The deadline that all workers who registered for the task have to submit their final results |
| Award (P) | Monetary prize (Dollars) in the task description. Range: $(0, \infty)$ |
| # Registration (R) | The number of registrants that are willing to compete on the total number of tasks in a specific period of time. Range: $(0, \infty)$ |
| # Submissions (S) | The number of submissions that a task receives by its submission deadline in a specific period of time. Range: (0, #registrants] |
| Technology | Required programing language to perform the task |
| Platform | The associate platform that a task is performing in that |
| Task type | Type of challenge depends on the development phase |
| Task Status | Completed or failed tasks |

### 3.3 Dataset Preparation

After data cleaning and removing tasks that were canceled per requestors wish, the number of tasks reduced to 4262 tasks. Because it is our interest to study task similarity effectiveness on task competition level and task stability level, we calculate, the number of similar arrival tasks per day as well as the number of open similar task in the platform per day. Then we clustered similar tasks to four groups of 60% task similarity, 70% task similarity, 80% task similarity, and 90% task similarity.

### 3.4 Empirical Studies and Design

Three analysis are designed based on the above metrics and proposed research questions, figure 1. Specifically, we are interested in investigating the following analysis in CSD:

**RQ1 (Task Similarity Level):**
*Task Similarity Analysis:* To analyze task similarity in the platform there is a need to understand the tasks' local distance from each other and task similarity factor based on it.

*Def. 1:* Task local distance ($D_i s_j$) is a tuple of all tasks' attributes in the data set. In respect to introduce variables in table 1, task local distance is:

$$D_{i}s_{j} = (\text{Award, Registration date, Submission Date, Task type, Technology, Task requirement}) \quad (1)$$

*Def.2:* Task Similarity Factor ($TS_{i,j}$) is a dot product and magnitude of the local distance of two tasks:

$$TS_{i,j} = \frac{\sum_{i,j=0}^{n} Disi(Tj,Ti)}{\sum_{i,=0}^{n} \sqrt{Disi(Ti)} * \sum_{j=0}^{n} \sqrt{Djsj(Tj)}} \quad (2)$$

*Similar Task Arrival:* It is important to measure the number of similar arrival tasks as well as similar available tasks per day per newly arrived task. Then we will analyze the correlation between task status, and similar arrival task per task similarity cluster in the dataset. Therefore, there is a need to understand task failure and task success per similarity cluster.

*Def.3:* Task Failure ($TF_j$) per similarity cluster is the number of failed tasks ($ft_i$) in the platform, which arrived at the same calendar day in the same similarity cluster:

$$TF_j = \sum FT_i \quad (3)$$

Where; $TS_{i,j} >= 0.9, 0.8, 0.7, 0.6$

*Def.4:* Task Success ($TSu_j$) per similarity cluster is the number of completed tasks ($ct_i$) in the platform, which arrived at the same calendar day in the same similarity cluster:

$$TSu_j = \sum ct_i \quad (4)$$

Where; $TS_{i,j} >= 0.9, 0.8, 0.7, 0.6$

In this part, we will be analyzing the impact of available similar tasks on task completing the level and the correlation of different degrees of task similarity on each other.

**RQ2 (Task Competition Level):** Workers' response on the same tasks in comparison with similar tasks will be analyzed in order to understand the task competition level to the same group of similar tasks.

*Def.5:* Task Competition Level, TCL (i,k), measures average registration for task i from registered similar arrival tasks.

$$TCL_{i,k} = \sum_{k=1}^{n}(TRi)/n \quad (5)$$

Where; $TSi, j \geq 0.6, 0.7, 0.8, 0.9$

**RQ3 (Task Stability Level):** Average valid submissions for task i in the same group of similar tasks illustrate task stability to be completed in the platform.

*Def.6:* Tsk Stability Level, TSL(i), measures average valid submissions of open tasks in workers queue in the same period of time that worker i take task j.

$$\text{TSL}_i = \sum_{k=1}^{n} S(i) \ / \ n \tag{6}$$

Where ; $TSi, j \geq 0.6, 0.7, 0.8, 0.9$

## 4 Empirical Result

### 4.1 Task Similarity Level (RQ1)

Our analysis showed that on average 76 new tasks per week were posted on the platform. On average 2 tasks canceled by task owners' requests, 1 faced zero registration and 8 failed based on zero submission or non-qualified submissions. Such observation means 14% of task failure in the platform. We further looked at the task failure distribution in the platform based on the level of task similarity. Figure 2 presents the trend of task arrival per task similarity group. As it is clear all the task groups are following an increasing trend. While the group of tasks with the similarity of 60% and 70% is more centered around the mean, 80%, and 90% similarity are skewed towards lower than mean. The lower level of similarity provides higher diversity in terms of available tasks. This means that the pool of tasks with the similarity of 60% is providing a higher level of choice for workers to switch context and compete on new tasks. Also, it seems the lower similarity among tasks leads to a higher level of completion among available tasks in the platform.

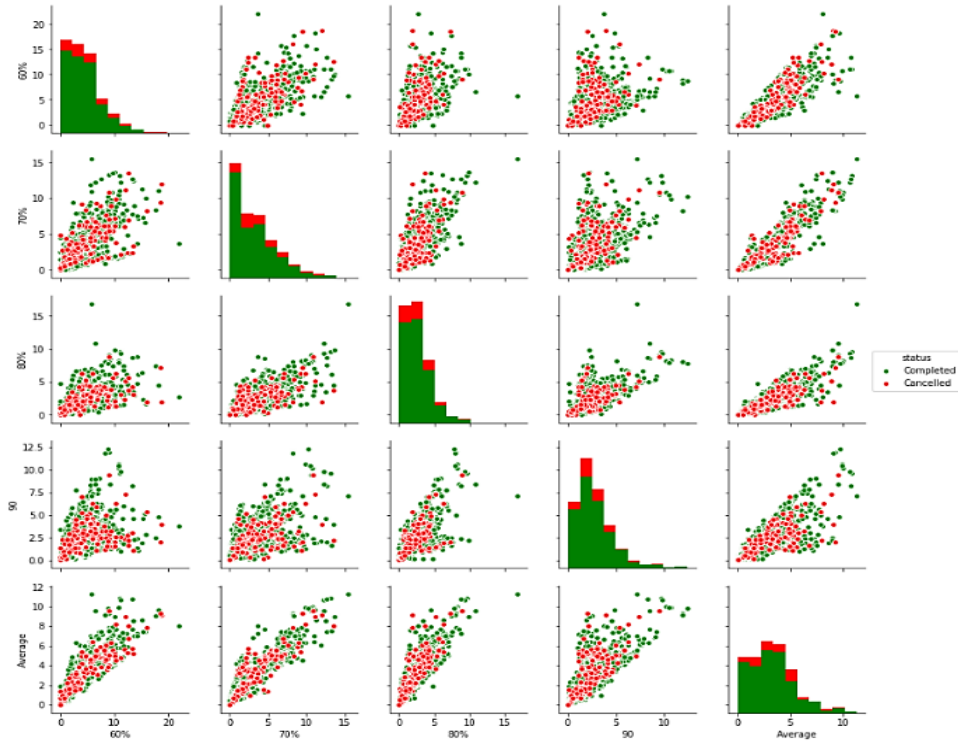

**Fig. 2.** Trend of task Success per Task Similarity Group

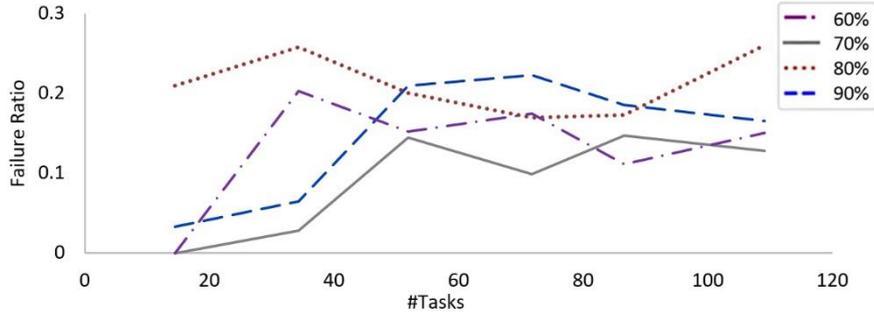

**Fig. 3.** Ratio of arrival similar task in the platform

Figure 3 illustrates the average failure ratio of tasks with respect to similar tasks arrival in the platform per week in four different similarity segments. As it presents, by increasing the task similarity level in the pool of available tasks, failure ratio increases. In the segment of 60% task similarity, failure ratio increases from 0% for less than 20 available tasks to 15% for more than 100 similar tasks. This ratio increased from 0% to almost 13% for the segment of 70% similarity. In the 90%, the similarity failure ratio increased from 4% for less than 20 tasks to 16% for more than 100 tasks. Interestingly, in the segment of 80% tasks, failure ratio was on average 20%. However, it dropped to around 17% for tasks in the range of 60 and 80, it again increased to 20% for more than 100 similar tasks available.

*Finding 1.1:* Higher level of similarity among available tasks in the platform leads to a higher level of task failure ratio.

### 4.2 Task Competition Level (RQ2)

In order to a better understanding of the influence of task similarity on workers' attraction and consequently competition level, we studied the distribution of competition level of tasks in the different groups of task similarity. Figure 4 illustrates the probability density of competition level under each similarity group.

As it is clear in figure 4, tasks with the similarity of 90% attract a higher level of competition with bio-model probability. The highest probability of attracting workers to the competition level is 16% with an average of 190 workers and the second pick of attracting workers is 0.6% with an average of 280 workers. interestingly 90% task similarity cannot attract more than 580 workers per task. 80% task similarity is a tri-model with the highest probability of 11% and an average competition level of 300 workers, and the minimum probability of 0.2% with an average competition level of 100 workers. This category of tasks cannot attract more than 500 workers. 70% task similarity is a tri-model that would drop the probability of attracting workers in the competition to 0.4% with an average of 430 workers. Tasks with the similarity level of 70% can attract up to 700 workers. Interestingly, the 60% task similarity dropped competition level to 0.56% with an average of 180 workers and maximum of 980 workers.

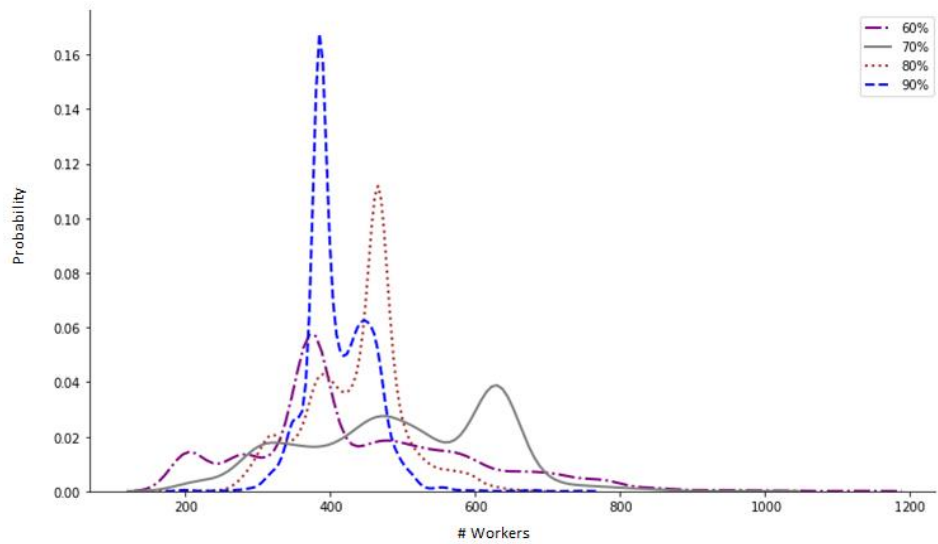

**Fig. 4.** Probability Density of Task Competition Level per Similarity Group

*Finding 2.1:* Higher level of similarity among available tasks in the platform (i.e 90% and 80% task similarity) leads to lower probability receiving registration per competition.

*Finding 2.2:* Lower level of similarity among available tasks in the platform (i.e 60% task similarity) leads to a higher probability of receiving higher registration level per competition.

### 4.3   Task Stability Level (RQ3)

In order Task Stability level is another important factor in task outcome besides competition level. Therefore, we investigated a layer deeper and analyzed receiving valid submissions per task based on the open tasks in a different cluster of task similarity. Figure 5 illustrates the distribution probability of receiving valid submissions per task similarity group. As it is shown in figure 5, receiving valid submissions by tasks follows the polynomial distribution. Table 2 summarized the statistical analytics of the distribution of valid submissions ratio per task similarity group.

Table 2. Summary of Valid Submission Ratio

| Metrics | 90% Valid Sub | 80% Valid Sub | 70% Valid Sub | 60% Valid Sub |
|---|---|---|---|---|
| Average | 27.00 | 23.00 | 31.00 | 27.00 |
| Min | 24.00 | 10.00 | 22.00 | 15.00 |
| Max | 30.00 | 28.00 | 35.00 | 33.00 |
| Median | 28.00 | 26.00 | 33.00 | 30.00 |
| Stdev | 2.00 | 6.00 | 5.00 | 7.00 |

The probability of a task receives valid submission in the pool of tasks with the similarity of 90% is 5%. This means 350 tasks receive valid submissions. While the probability of receiving valid submissions for tasks with the similarity of 80% increased up to 7%, only 180 tasks in this group receive a valid submission. The cluster of tasks with the similarity of 70%, receiving valid task submission provides a tri-model in which two of them provides the probability of 4% with an average of 110 and 300 valid submissions respectively, and the third one provides the probability of 2% with an average of 230 valid submissions. Interestingly, tasks with a similarity of 60% lead to more than 20% probability of receiving valid submissions with an average of 110 valid submissions.

*Finding 3.1:* Lower level of similarity (i.e 60%) among available tasks in the platform lead to a higher probability of receiving task stability.

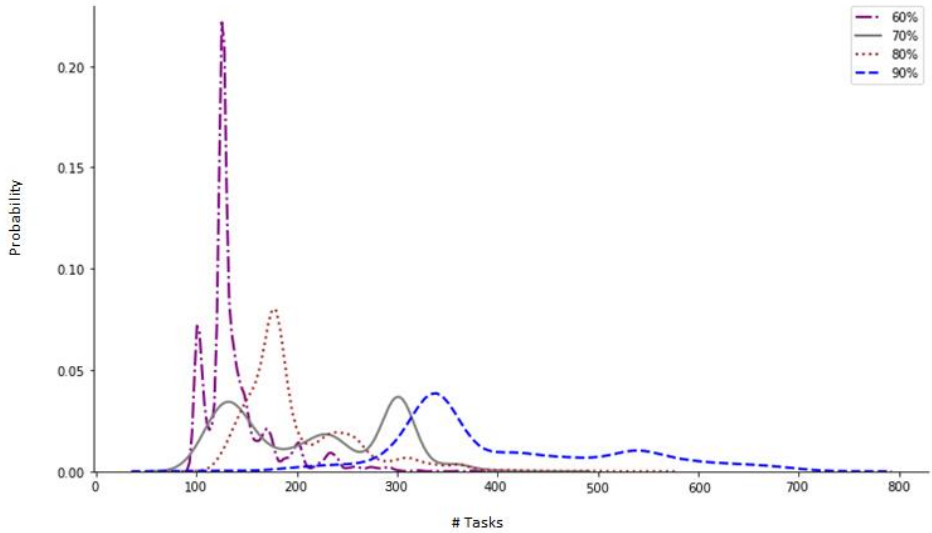

**Fig. 5.** Probability Density of receiving valid submission

# 5 Discussion and Evaluation

## 5.1 Similar Task Arrival

To successfully crowdsourced a software project in a crowdsourcing platform, besides understanding the task dependency in a project, it is important to identify the pattern of similar available tasks in the platform. The pool of available similar tasks would highly impact task competition level, task stability level and consequently project success. Also, facing a higher number of similar arrival tasks at the same time doesn't necessarily mean a bigger pool of available similar tasks and less chance of attracting qualified workers in the platform.

As it is shown in figure 1, arriving more than 100 available similar tasks in the platform will increase the chance of task failure up to 15%. Moreover, increasing task similarity ratio from 60% to 90% directly impacted on task failure. The increasing degree of task similarity in the platform would ease switching context for workers and lead to lots of failure tasks due to starvation or zero submissions [7,30].

Our research showed that having multiple choices for workers will negatively impact on competition level and receiving qualified submissions, finding 1.1.

## 5.2 Task Competition Level

To assure of having a successful project in CSD, not only it is valuable to have a higher competition level to attract enough available qualified workers, but also it is important to have a good worker elasticity among different tasks. This fact makes the importance of knowledge workers' task registration patterns. Our study showed that generally a long line of active registered tasks per worker is negatively correlated with their submissions ratio. Besides having a high number of non-similar tasks in workers' basket indicates attracting a not qualified worker for the tasks, finding 2.1.

Moreover, finding 2.2 showed that, although a longer line of non-similar tasks in a worker's queue is not good, having a long line of similar tasks in workers' queue may lead to less worker elasticity[10]. This makes workers not be available to take a series of tasks of the same project or be very busy working on another similar and perhaps more attractive tasks and ignore other registered tasks.

However, 75% of workers retook the repeated task, and the project didn't face a high level of resource mismatching, team elasticity was low. Therefore, a bigger pool of open similar tasks in the platform is not always a good strategy for attracting qualified workers with high elasticity.

## 5.3 Task Stability Level

Our result presented that a higher number of available similar tasks means a higher number of different choices for a worker and a lower chance of receiving qualified submissions for a task. According to finding 3.1 workers with the lower queue of tasks are more reliable to retake task from the same project. This fact can be beneficial due to having some background and information about the project.

Since the different choice of task taking for workers will directly impact workers' availability to work on tasks, the higher queue of tasks generally and the queue of similar tasks especially represents a lower chance of receiving qualified submissions [30], and consequently means a higher chance of resource discrepancy. In order to overcome this issue not only a project manager should track open similar tasks in the platform upon task arrival, but it is also required to track workers' performance and activity in the platform in the same frame time.

### 5.4 Task Status Prediction

As finding 1.1 reports, the level of task similarity in the CSD platform directly impacts on task status. In order to prevent task failure as much as possible, a CSD project manager needs to predict the best time for posting the task. Therefore, in order to study the task failure level in the CSD platform, we build a prediction model using a time series analysis based on the dataset we used in this research. To build the model we used task arrival date, task similarity level with available tasks, task competition level and task submission level in the CSD platform and task status as the outcome of the prediction model. We have selected 4 configurations of task similarity i.e 60%, 70%, 80%, and 90% respectively for the 30days of arriving tasks based on different similarity levels. For building and evaluating of the model we created two sets of the training set and sample set of data. The training set contains information required for a task where a worker actually registered and made a submission for the task earlier than the submissions date. Similarly, the testing set provides the same information but for all the tasks that were submitted on the deadline. The prophet library introduced by Facebook [32] was used for building and evaluating the predictive time series model.

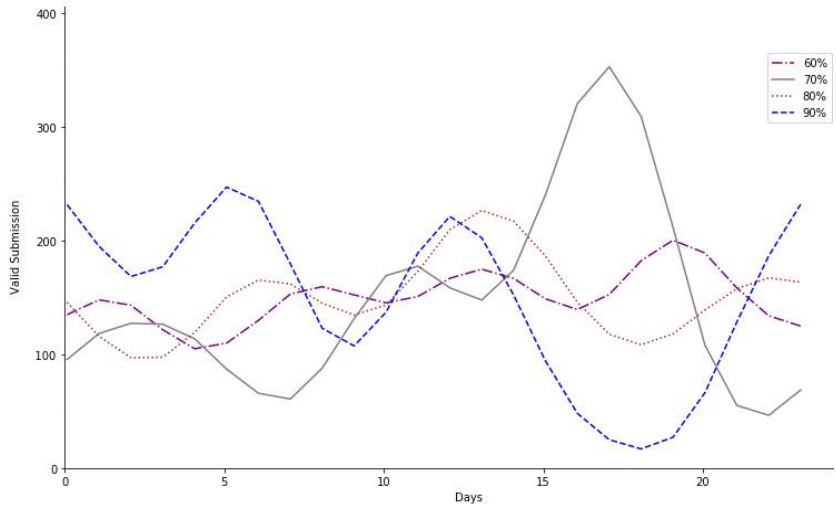

**Fig. 6.** Prediction of Valid Submission Per Task Similarity Group

Figure 6 illustrates the result of task valid submission prediction in the platform. As it is shown tasks with 60% similarity have a stationary prediction for making a submission, with an average of 135 submissions and the submission increased by passing time. Interestingly, while tasks with 70% and 90% level of similarity will not receiving a promising trend of receiving valid submissions, while tasks with 80% similarity are receiving valid submissions with an average of 180.

However, tasks with 60% similarity continue delivering valid submissions close to mean, tasks with 80% similarity experience a higher level of variation. It seems the best time for arriving a new task in the platform in order to have the highest chance of receiving valid submission is when there is the highest level of tasks with 60% similarity available in the pool of open tasks.

### 5.5 Limitations

First, the study only focuses on competitive CSD tasks on the Topcoder platform. Many more platforms do exist, and even though the results achieved are based on a comprehensive set of about 5000 development tasks, the results cannot be claimed externally valid. There is no guarantee the same results would remain exactly the same in other CSD platforms.

Second, there are many different factors that may influence task similarity and workers' decisions in task selection and completion. Our similarity algorithm is based on known task attributes in Top coder. Different similarity algorithms may lead us to different but almost similar results.

Third, the result is based on task similarity only. Workers' network and communication were not considered in this research. In the future, we need to add this level of research to the existing one.

## 6 Conclusions

Task preparation is one of the most important issues in the crowdsourcing development world. Crowdsourced tasks perform by an unknown group of workers, who choose tasks based on personal and mostly unknown utility factors. This issue raises the concern about planning a task arrival on the platform. Task similarity impacts on task arrival indirectly. Our research shows that not only a higher level of similarity among available tasks in the platform negatively effects on task success but also the number of arrival similar tasks in the platform would impact competition level per task as well as workers' elasticity. Moreover, workers' queue of task impacts workers' discrepancy and their willingness to continue working on the same project's tasks.

This paper reports that 1) higher number of arrival tasks with similarity level greater than 70% will negatively impact on task competition level, 2) Bigger pool of similar open and arrival tasks would lead to lower worker attraction and elasticity, and 3) Workers who register for tasks with lower similarity level are more reliable to make a valid submission and 4) arriving task to the pool of 60% similar task will provide the highest chance of receiving valid submission.

In future, we would like to focus on the similar crowd worker behavior and performance based on task similarity level and try to analyze workers trust network as well as a task-worker performance network to report more decision elements according to task size and date of uploading, achievement level, task utilization, and workers performance.